\documentclass[twocolumn,aps,superscriptaddress,prl]{revtex4}
\usepackage{graphicx}
\usepackage{times}
\usepackage{amsmath}
\usepackage{amssymb}
\usepackage{units}
\usepackage{stmaryrd} 

\begin{document}
\preprint{0}

\title{A chemical imaging and Nano-ARPES study of well-ordered thermally reduced SrTiO$_{3}$(100)}

\author{Emmanouil Frantzeskakis}

\address{Synchrotron SOLEIL, L'Orme des Merisiers, Saint Aubin-BP 48, 91192 Gif sur Yvette Cedex, France}

\author{Jos\'{e} Avila}
\address{Synchrotron SOLEIL, L'Orme des Merisiers, Saint Aubin-BP 48, 91192 Gif sur Yvette Cedex, France}

\author{Maria C. Asensio}
\address{Synchrotron SOLEIL, L'Orme des Merisiers, Saint Aubin-BP 48, 91192 Gif sur Yvette Cedex, France}
\date{\today}

\begin{abstract}

The structural and electronic properties of thermally reduced SrTiO$_{3}$(100) single crystals have been investigated using a probe with real- and reciprocal-space sensitivity: a synchrotron radiation microsopic setup which offers the possibility of Scanning Photoemission Microscopy and Angle Resolved Photoelectron Spectroscopy (ARPES) down to the nanometric scale. For the first time, we have spectroscopically imaged the chemical composition of samples presenting reproducible and suitable low-energy diffraction patterns after following well-established thermal reduction protocols. At the micrometric scale, Ca-rich areas have been directly imaged using high-energy resolution core level photoemission. Moreover, we have monitored the effect of Ca segregation on different features of the SrTiO$_{3}$(100) electronic band structure, measuring ARPES inside, outside and at the interface of surface inhomogeneities with the Ca-rich identified areas. In particular, the interaction of Ca with the well-known intragap localized state, previously attributed to oxygen vacancies, has been investigated. Moreover, the combination of direct imaging and spectroscopic techniques with high spatial resolution has clarified the long-standing dilemma related to the bulk or surface character of Ca segregation in SrTiO$_{3}$. Our results present solid evidence that the penetration depth of Ca segregation is very small. In contrast to what has been previously proposed, the origin of the long-range surface reconstructions can unlikely be associated to Ca due to strong local variations of its surface concentration. 

\end{abstract}

\maketitle

\section{I. Introduction}

Strontium titanate (SrTiO$_{3}$ or simply STO) has attracted enormous scientific interest due to its numerous applications. STO surfaces
have been used as substrates for the growth of 2D high-T$_{\textmd{c}}$ superconductors \cite{Sum1995,Chaudhari1987,Endo2003}, metallic
thin films \cite{Kawasaki1994,Koslowski2002,Vlachos2004} and nanocrystals \cite{Silly2005,SillyAPL2005,Silly2006,Fu2002}. STO enjoys remarkable photoelectrolytic
\cite{Mavroides1976} and optical properties \cite{Kan2005}, while it has been used as a template for oxide interfaces which could be exploited in all-oxide devices
\cite{Ohtomo2004,Takagi2010,Thiel2006}. Moreover, the recent discovery of a high-mobility two-dimensional electron gas (2DEG) on its (100) surface has motivated many studies of its electronic structure
\cite{Meevasana2011,Santander2011,Chang2010}. The number of surface reconstructions of the widely studied STO(100) surface is as rich as the aforementioned applications range.
Previously observed reconstructions comprise: $1\times1$ \cite{Cord1985,Naito1994,Jiang1995}, $2\times1$ \cite{Cord1985,Naito1994,Jiang1995,Erdman2002,Castell2002,Johnston2004}
, $2\times2$ \cite{Andersen1990,Moller1999,Jiang1995,Nishimura1999,Ohsawa2010,SillySS2006}, $4\times4$ \cite{Kubo2003}, c$(4\times2)$ \cite{Jiang1999,Castell2002-1}, c$(4\times4)$ \cite{Castell2002-2}, c$(6\times2)$ \cite{Jiang1999,Lanier2007},
$\sqrt{13}\times\sqrt{13}\textmd{R}33.7^{\textmd{o}}$ \cite{Naito1994,Kienzle2011} and $\sqrt{5}\times\sqrt{5}\textmd{R}26.6^{\textmd{o}}$ \cite{Newell2007,Gonzalez2000,Kubo2001,Tanaka1993}.

Several cleaning procedures have been established to study the morphology and the electronic structure of STO(100). Most previous studies were
performed either in cleaved \cite{Aiura2002,Santander2011,Meevasana2011} or UHV annealed samples \cite{Liang1995,Gonzalez2000,Jiang1999,Takizawa2009,Chang2010}, albeit there is
no general consensus on the best preparation method. Ca segregation is a 
long withstanding problem encountered even after following well-established cleaning protocols \cite{Andersen1990,Polli1999,Herger2007}. 
Atomic Force Microscopy images have investigated the segregation of Ca impurities on the STO(100) surface but they are not sensitive to the penetration depth of the contaminants \cite{Polli1999}. 
Wet chemical etching has been reported to diminish the amount of
Ca impurities \cite{Polli1999,Herger2007,Kawasaki1994,Ohnishi2004}. Nevertheless, the number of necessary etching cycles is under debate and on the other hand such a
cleaning procedure may result in the segregation of different contaminants \cite{Herger2007}. Apart from STO, Ca segregation has been reported on the 
surface of different oxide materials \cite{Gajdardziska1993,Fukui1999,Mariotto2004,Zhang1998}.

We reinvestigate the effect of Ca segregation on the STO(100) surface using a novel experimental tool: the experimental setup of the ANTARES beamline at SOLEIL,
which offers Scanning PhotoEmission Microscopy (SPEM) and Angle Resolved PhotoElectron Spectroscopy (ARPES) down to the nanometric scale.
The scope of our work is threefold. First, we establish the micrometric favorable sites for the segregation of Ca impurities and we demonstrate the existence of Ca-free areas.
Secondly, we report on the effect of Ca segregation on the electronic structure of STO(100) by monitoring the spatial variations  of the k-resolved valence band (VB) and of the intragap
localized surface state. Last but not least, we show the capabilities of a state-of-the-art experimental tool which provides very high reciprocal- and real-space sensitivity. As a result, not only
are we able to directly image the in-plane segregation of Ca but we can also draw conclusions about the bulk concentration of such impurities.

\section{II. Experimental Details}

STO(100) single crystals (Crystal GmbH and SurfaceNet GmbH) were mounted on resistive Si heaters and cleaned by thermal annealing in 
UHV conditions. Each sample was sequentially heated up to $1200^{\textmd{o}}$C where it was kept for $20$ minutes. At all 
stages of the cleaning procedure the pressure was lower than $10^{-9}$ mbar. All sample surfaces passed through a series of reconstructions until they finally exhibited a sharp LEED pattern characteristic of a $\sqrt{5}\times\sqrt{5}\textmd{R}26.6^{\textmd{o}}$ arrangement (Fig. 2). 

SPEM and (AR)PES measurements were conducted at the 
ANTARES beamline of SOLEIL. The ultimate angle and energy resolutions
of the ARPES apparatus (Scienta R4000) are better than $2$ meV and $0.1^{\textmd{o}}$, respectively.
The former was set to $30$ mev by the conditions of the described experiment. 
The sample temperature during measurements was $120$K.

The experimental endstation facilities are schematically depicted in Fig. 1. The incoming photon beam can be 
focused on a nanometric spot by means of a Fresnel Zone Plate (ZP), while an Order Sorting Aperture (OSA) removes undesirable higher-order 
diffraction effects. The sample is mounted on a high-precision scanning stage with five degrees of freedom. Piezo-scanners allow zero 
backlash and nanometric accuracy in the linear motion. There is an interferometric control of the sample and ZP positions. Images of the sample chemical composition and angle resolved valence band states can be obtained by the synchronized scanning of the sample 
with respect to the focused photon beam, while the electron analyzer collects photoelectrons with selected kinetic energy. This is the most direct 
approach to acquire spatially-sensitive ARPES spectra at a nanometric scale (i.e. nano-ARPES). The same experimental setup
offers the possibility of conventional spatially-integrated (AR)PES measurements. In this case, the photon beam is focused directly on the sample without the use of 
a Fresnel Zone Plate. The approximate spot size with (without) ZP under the described experimental conditions is $100$ nm ($100$ $\mu m$). Further details about the ANTARES experimental endstation will be given elsewhere.
\begin{figure}
  \centering
  \includegraphics[width = 8.7 cm]{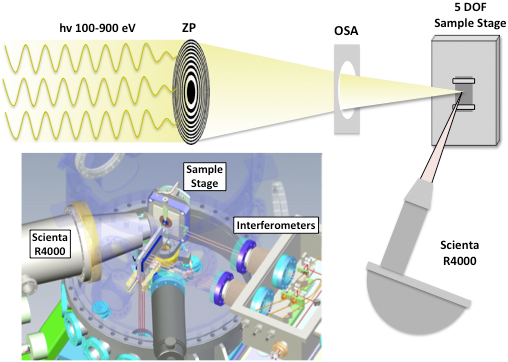}
  \caption{(color online)
A schematic representation of the experimental setup consisting of the Fresnel Zone Plate (ZP), the Order Sorting Aperture (OSA), a high-precision sample manipulator 
for angular and linear motion (five Degrees Of Freedom) and the electron analyzer. The inset presents a technical drawing of the setup including the interferometric control
of the sample and ZP positions.}
\label{fig1}
\end{figure}\begin{figure}[!b]
  \centering
  \includegraphics[width = 8.7 cm]{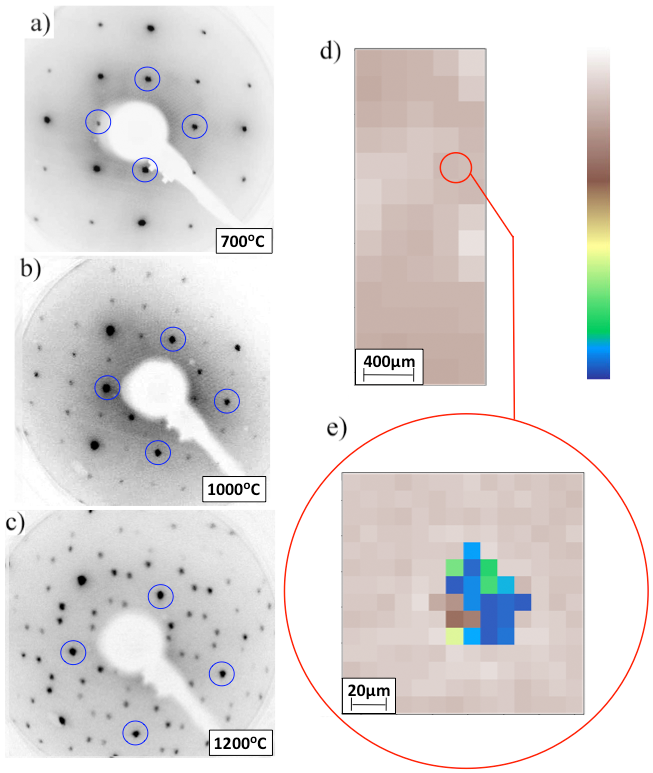}
  \caption{(color online)
  (a)-(c) The sequence of LEED patterns with increasing annealing temperature. Surface reconstruction changes from $1\times1$ (a) to $2\times2$ (b) and $\sqrt{5}\times\sqrt{5}\textmd{R}26.6^{\textmd{o}}$ (c).
  Circles mark the $1\times1$ spots of the reciprocal unit cell. (d) A spatial image of the sample following the integrated intensity of the Sr $4p$ peak
and using the conventional PES facilities of the experimental endstation. (e) A higher resolution spatial image of (c) using the Zone Plate focusing of the incoming photon beam. The
scanned area of (d) which is magnified in (e) is marked by a circle. Photoemission intensity follows the attached color scale, where signal-to-noise ratio increases from the bottom to the top. The photon energy is $100$ eV.}
\label{fig2}
\end{figure}
  
\section{III. Results}
  
The right panels of Fig. 2 illustrate the advantages of the ANTARES microscope with respect to a conventional (AR)PES setup. 
Fig. 2(d) is a spatial image of a thermally reduced STO(100) single crystal surface by following the integrated intensity of the Sr $4p$ PES peak while scanning the 
sample in front of a conventional photon beam (i.e. no ZP). The observed intensity variation is negligible. When the focusing 
capabilities of the ZP are used to zoom into different areas of the sample, one may observe clear features as presented in Fig. 2(e).
These can be attributed directly to chemical inhomogeneities (different Sr concentration). These features,  due to the lack of high lateral resolution, cannot be identified neither by conventional PES [Fig. 2(d)], nor by the LEED image [Fig. 2(c)].

In Fig. 3(a) we present a larger spatial scan evidencing that two such inhomogeneities can be observed within an area of $500 \times 500$ $\mu$m$^{2}$. 
As in Fig. 2(e), the image has been acquired following the integrated intensity of the Sr $4p$ peak. 
A unique advantage of the ANTARES setup is that the imaging capabilities are extended in the UV range down to $100$ eV. 
Therefore, as a complement to the core-level imaging, one may obtain information about the spatial variation of electronic states with low binding energy. 
Fig. 3(b) follows the intensity of the STO VB in the same area imaged in Fig. 3(a). Similar intensity variations are observed at the same sample positions 
proving that VB is a sensitive probe to image surface inhomogeneities.
\begin{figure}
  \centering
  \includegraphics[width = 8.7 cm]{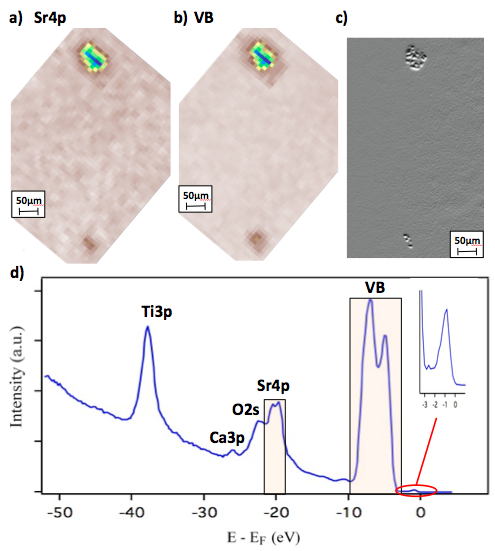}
  \caption{(color online)
A spatial image of the sample following the integrated intensity of (a) the Sr $4p$ peak
and (b) the VB using the ZP focusing of the incoming photon beam.
(c) An optical micrograph of the same sample area revealing the existence of surface defects.
(d) An angle-integrated PES spectrum evidencing Ca impurities. The inset is a magnified
version of the intragap localized surface state. The energy regions used in (a) and (b) for the
acquisition of spatial images are marked in the PES spectrum. Photoemission intensity follows the color scale of Fig. 2, where signal-to-noise ratio increases from the bottom to the top. The photon energy is $100$ eV.}
\label{fig3}
\end{figure}
\begin{figure}[!b]
  \centering
  \includegraphics[width = 8.6 cm]{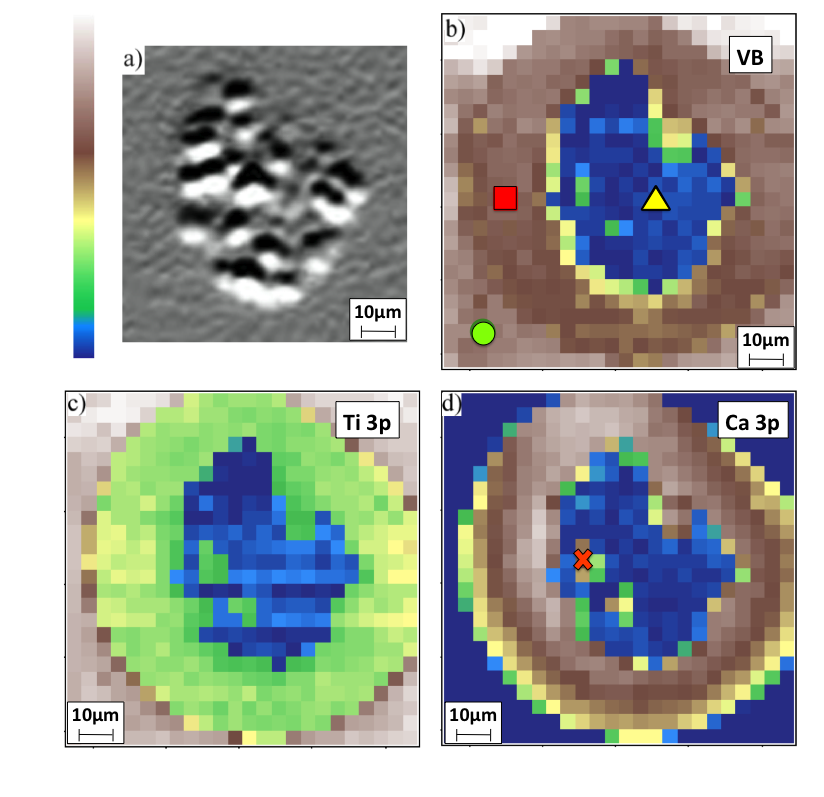}
  \caption{(color online)
Spatial images (zoom) of the upper inhomogeneity observed in Fig. 3: (a) Optical micrograph.
[(b)-(d)] Integrated intensity of (b) the VB, (c) the Ti $3p$ peak and (d) the Ca $3p$ peak
using the ZP focusing of the incoming photon beam. Ca impurities decorate the surface defect
adding a ring-like contour with an approximate width of $15\mu$$m$. Marks in (b) denote the exact positions of the (AR)PES
spectra presented in Figs. 5 and 6.
Photoemission intensity follows the attached color scale, where signal-to-noise ratio increases from the bottom to the top.
The photon energy is $100$ eV.}
\label{fig4}
\end{figure}
    
Fig. 3(c) presents an optical micrograph of the same sample area, also after UHV annealing. The inhomogeneities are visible and can be 
attributed to surface defects on top of the sample. Their presence was verified on different STO(100) samples. Fig. 3(d) presents an angle-integrated PES spectrum 
on the same STO(100) single crystal using a conventional photon beam (i.e. no ZP). Due to the thermal reduction of the sample, a state arises within the electronic bandgap.
It has been previously attributed to the formation of oxygen vacancies  \cite{Gonzalez2000,Santander2011,Aiura2002}. One may also notice the appearance of a small Ca-related peak, one of the already reported impurities 
in STO(100) single crystal surfaces \cite{Andersen1990,Polli1999,Herger2007}. In the following, we will use the experimental capabilities of the ANTARES setup to investigate the origin of the aforementioned defects, determine 
their connection to Ca impurities and comment on the modifications of the electronic band structure.

Figures 4(b)-4(d) are higher-resolution microscopic images of the upper inhomogeneity observed in Figures 3(a) and 3(b). They have been 
acquired using various PES intensity peaks as experimental probes. Fig. 4(a) presents an optical micrograph of the 
same feature, as evidenced by the resemblance of the contours of the two images [e.g. 4(a) vs. 4(b)]. One can readily distinguish two different areas on the microscopic 
images of the inhomogeneity; (i) the defect itself as observed by the optical microscope and (ii) a ``ring" of around 15 $\mu m$ which spreads around the contour of the defect. 
The optical micrograph is insensitive to the ring.
  
The results presented in Fig. 4 have been acquired by following the integrated intensity of the VB [Fig. 4(b)], the Ti $3p$ [Fig. 4(c)] and the Ca $3p$ [Fig. 4(d)] PES peaks. 
The distance of the ZP and the sample has been optimized both inside and outside of the inhomogeneity, while the integrated photoemission intensity at different places has 
been normalized by the corresponding background intensity. In this way, we have ensured that the observed intensity variations are not related to the ZP focalization but they instead arise 
from real modifications of the electronic structure. Fig. 4(d) suggests that the amount of Ca on the ring is increased with respect to other areas. This is also captured by the PES spectra of Fig. 5(d) 
which are acquired with a nanospot beam at two different points of the sample; i.e. the lower spectra outside the ring and the defect and the 
upper spectra on the ring [the exact positions are marked by a circle and a square in Fig. 4(b)]. The spatial variation of different Ca PES peaks is followed. 
There is a substantial increase of Ca concentration around the defects in a ring-like area with a thickness of $15$ $\mu m$. 
Within a few $\mu m$ away from this area, the intensity lineshapes change to the lower spectra of Fig. 5(d)
suggesting that these are Ca-free areas. Moreover, in the interior of the observed inhomogeneity there are only 
trace amounts of Ca [lower spectrum in Fig. 5(e)], localized in areas which yield a similar Ca $3p$ intensity as on the ring [e.g. one of them is marked by ``x" in Fig. 4(d)]. After scanning the surfaces of various samples we have verified that these abrupt intensity variations of Ca-related peaks cannot be observed by conventional PES. Such a spatially-integrated technique cannot detect the defective area and it would still require moving the photon beam far from all the observed inhomogeneities by a few hundreds of $\mu m$ in order to work on a Ca-free area. In other words, the PES spectrum of Fig. 3(d) presents slight modifications only after scanning a wide spatial range, thus forbidding an exact determination of Ca-free areas, as it is possible with our spatially-sensitive setup.
\begin{figure}
  \centering
  \includegraphics[width = 8.4 cm]{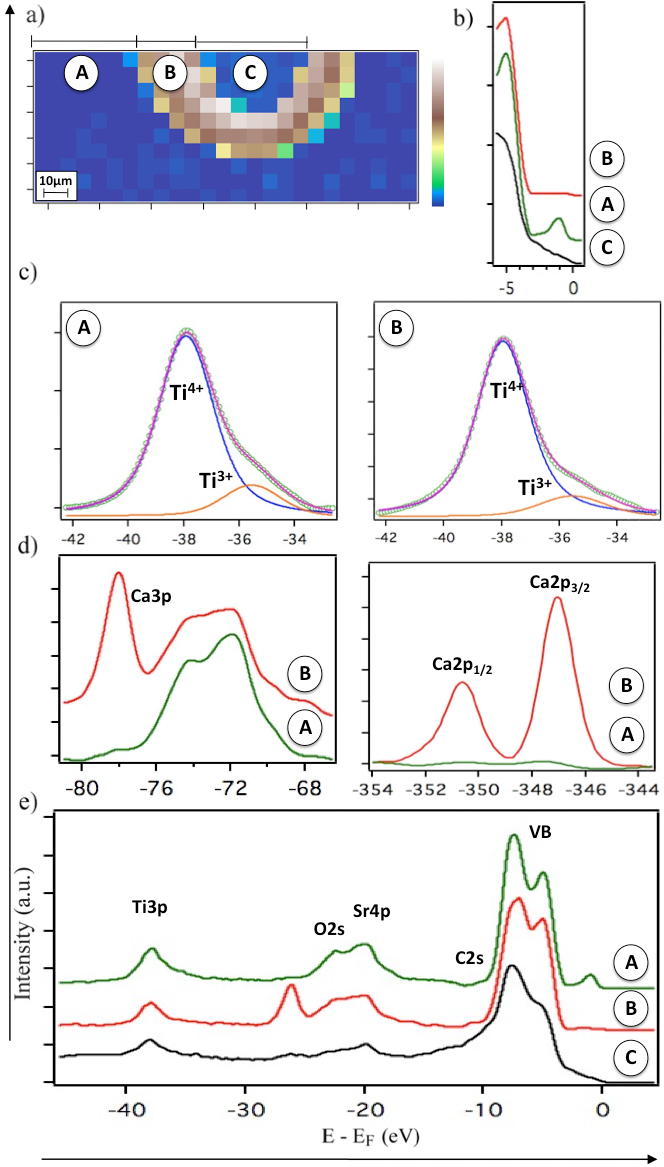}
  \caption{(color online)
  (a) A spatial image of a sample inhomogeneity following the integrated intensity of the Ca $3p$ peak with the ZP focusing. 
  ``A", ``B" and ``C" mark respectively the Ca-free region, the ring with high Ca segregation
  and the defective area. Photoemission intensity follows the attached color scale where signal-to-noise ratio increases from the left to the right. (b)-(e) Angle-integrated PES spectra at the different energy ranges
  shown in (a) acquired with the ZP focusing. Spectra were acquired at the positions marked in Fig. 4(b) and were reproduced for different points of the corresponding regions.
  (b) PES spectra near $E_{F}$ at regions ``B" (top), ``A"  (middle) and ``C" (bottom). Clear variations are observed in the intragap
  state. (c) The Ti $3p$ PES peak in regions ``A" (left) and ``B" (right). Lineshapes have been fitted with two Voigt functions where the total fit (thin solid line) reproduces the sequence of experimental points
  (dots). The peak at low binding energy corresponds to Ti$^{3+}$ and is a signature of oxygen depletion. In agreement with (b), the low binding energy peak is enhanced in region ``A". (d) PES
  spectra of the Ca $3p$ (left) and Ca $2p$ (right) peaks in regions ``A" (bottom spectra) and ``B" (top spectra). There is a large variation of Ca segregation within a few $\mu m$. (e) PES spectra acquired on the three
  different regions (i.e. ``A" to ``C" by moving from the top to the bottom) using the same energy range as in Fig. 3(d). Photon energies of $100$ eV and $700$ eV were used.}
\label{fig5}
\end{figure}
    
The integrated intensity of Ti $3p$ is decresing stepwise as one moves from a Ca-free area to the ring and finally to the defect [Fig. 4(c)].
Similar behavior is observed for the VB peak, albeit VB modifications are less sensitive to the ring [Fig. 4(b)]. The PES spectra are informative [Fig. 5(c)].
The asymmetric lineshape of Ti $3p$ is a signature of Ti$^{3+}$ electron states, i.e. the fingerprint of oxygen depletion. Ti$^{3+}$ contribution to the
total spectrum is enhanced in Ca-free areas [Fig. 5(c)-left] pointing towards a larger number of oxygen vacancies. In line with the above reasoning, oxygen vacancies
can be clearly identified in Ca-free areas by a large intragap peak near the Fermi level [Fig. 5(b)-middle]. This state is negligible on the Ca ring [Fig. 5(b)-top] and is replaced
by a structureless background at defective areas [Fig. 5(b)-bottom]. One should note that there is a substantial increase of intensity with respect to the corresponding intragap state
observed by conventional ARPES in the same sample area [Fig. 3(d)]. As a matter of fact, the spatially-integrated PES spectrum of Fig. 3(d) can 
be considered as the average of ``spatially-sensitive" PES spectra like those presented in Fig. 5(e). Within a few $\mu m$, segregation of Ca impurities and the existence of oxygen 
vacancies undergo related modifications, which cannot be identified with a conventional ARPES setup.
\begin{figure}
  \centering
  \includegraphics[width = 8.65 cm]{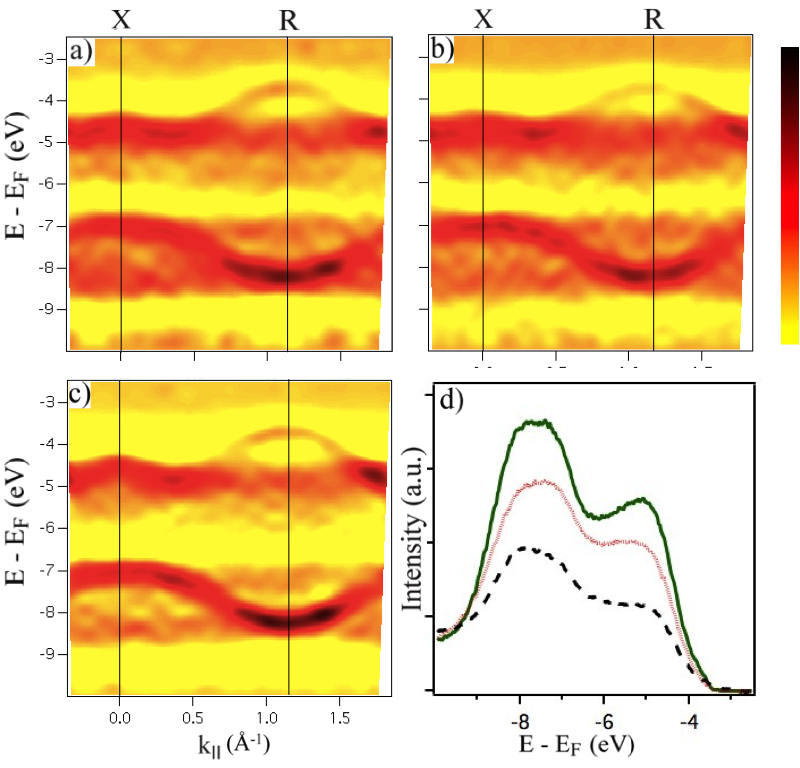}
  \caption{(color online)
VB dispersion along the X-R high-symmetry line of the Brillouin zone 
for three different positions on the sample: (a) on the surface defect, (b) on the ring from Ca segregated impurities, (c) outside the surface defect and $10$ $\mu m$ far from segregated Ca impurities.
Exact positions are marked in Fig. 4(b).
Due to the bulk character of the bands, there are no significant changes on the experimental band dispersion. The signal-to-noise ratio is strongly improved in (c). The second derivative of the photoemission intensity
has been used to enhance the experimental features. Photoemission intensity follows the attached color scale, where signal-to-noise ratio increases from the bottom to the top. 
(d) The angle-integrated PES spectrum of the VB corresponding to (a) (i.e. dashed line), to (b) (i.e. dotted line) and to (c) (i.e. solid line).
In defective and Ca-rich areas the VB lineshape becomes structureless and the background noise increases.}
\label{fig6}
\end{figure}
    
Fig. 6 presents electronic band structure diagrams acquired when the nanospot beam is localized at the three different points marked in Fig. 4(b). 
The intragap vacancy-related state is dispersionless \cite{Santander2011}, thus we have focused on the electronic dispersion of the VB. The results presented in Fig. 6 are a clear demonstration that the
experimental facilities allow data acquisition with sensitivity both at real and reciprocal space. The electronic band diagram of Fig. 6(c) has been obtained on a Ca-free area, while those of Fig. 6(a) and 6(b) 
have been respectively acquired in the interior of the defect and on the Ca-induced ring. The sample has been oriented along the [110] azimuth and the 
photon energy is $100$ eV. If one considers free-electron final states and an inner potential of $17$ eV \cite{Takizawa2009},
the $E-k$ diagrams are along the high symmetry X-R direction of the Brillouin zone. The underlying dispersion is in agreement with experimental 
and theoretical studies \cite{Takizawa2009,vanBenthem2001}. Comparing the three band diagrams, there are no major modifications in the overall dispersion of the bulk VB. This is a confirmation that the 
aforementioned defects but also the segregation of Ca impurities modify the surface of the STO single crystal but have no effect in the bulk. The bands of Fig. 6(c) are better resolved and 
have a more favorable signal-to-noise ratio in comparison to Fig. 6(a) and Fig. 6(b). As expected, Ca-free areas on the surface of STO(100) are advantageous in obtaining a high-quality 
electronic band structure by means of ARPES. These observations are confirmed by a comparison of the corresponding angle-integrated VB spectra in Fig. 6(d). 
Photoemission from points outside defects and Ca-segregation areas reveal a VB lineshape with clearer structure and decreased noise.

\section{IV. Conclusions}

As noted in Section I, Ca is a well-known impurity on the surface of STO(100) single crystals. On one hand, it diffuses from the bulk to the surface by high-temperature annealing \cite{Andersen1990,Herger2007}.
On the other hand, Ca$^{2+}$ ions adsorb to the surface of commercially prepared STO single crystals from polishing or post-polishing cleaning solutions \cite{Polli1999}. This last study suggested that
Ca impurities may segregate during heat treatment at surface defects such as steps. The data presented in Figs. 3 and 4 verifies the conjecture that Ca impurities decorate surface defects after thermal activation.
We have evidenced that there are strong variations of the local surface concentration of Ca. Therefore, in contrast to what has been previously proposed \cite{Andersen1990}, we assume that Ca is unlikely 
the origin of any surface reconstruction which would require a well-ordered distribution of impurities all over the surface.

The studied $\sqrt{5}\times\sqrt{5}\textmd{R}26.6^{\textmd{o}}$ reconstruction is a signature of a reduced environment. It has been previously attributed to the ordering of oxygen vacancies \cite{Gonzalez2000,Tanaka1993,Tanaka1994}
or to the phase separation of the surface region into a reconstructed Sr adlayer and TiO islands \cite{Newell2007,Kubo2001,Kubo2003}. In both cases oxygen depletion is a necessary requirement and is accompanied
by a substantial increase in the electrical conductivity \cite{Newell2007,Szot2002}. Moreover, the existence of oxygen vacancies on a SrTiO$_{3}$(100) has been proposed as the origin of the widely studied high-mobility 2DEG at the related LaAlO$_{3}$/SrTiO$_{3}$ interface \cite{Siemons2007,Kalabukhov2007,Ohtomo2004}. It presents an alternative scenario to the accommodation of a polar catastrophe \cite{Nakagawa2006}. In terms of the electronic structure, the conductivity increase and the related oxygen vacancies are accompanied by an intragap localized surface state as reported in
previous studies \cite{Gonzalez2000,Santander2011,Cord1985,Aiura2002,Tanaka1993} and shown in Fig. 3(d). Our work suggests that the intragap electronic state and hence the surface conductivity is
characterized by strong local variations intimately connected to the local concentration of Ca and to the existence of surface defects [Fig. 5(b)]. Ca segregation results into the complete extinction
of the localized state. On the other hand, Ca-free areas yield a well-defined peak accompanied by an enhanced core level signature of Ti$^{3+}$ electrons [Figs. 5(c)].

In contrast to the strong spatial variations of the intragap surface state, negligible modifications were observed on the dispersion of the VB. This result confirms that the studied defects and 
related Ca segregation modify the sample surface but have little effect on its bulk. Ca contamination is localized around the aforementioned defects but
has a negligible penetration depth. Such micrometric defective areas exist on samples acquired from different vendors and their formation may be attributed to
the final stages of the commercial preparation of a STO(100) single crystal. They are decorated by Ca impurities during UHV annealing, thus shrinking the defect-free area of the sample.
As evidenced in Fig. 6, ARPES measurements in a region far from Ca-segregated impurities and surface defects do not only ensure the existence of an intragap surface state on a well-ordered surface, but also yield
a diminished background, narrower lineshapes and enhanced photoemission intensity of the VB.

Apart from revealing the favorable micrometric segregation sites of Ca impurities and exploring their effect on the electronic band structure of STO(100), our work evidences the experimental capabilities
of a new setup. SPEM as performed at the ANTARES endstation can be extended down to $100$ eV allowing ARPES measurements with very high spatial sensitivity.
Fig. 4(d) captures the passage from a PES spectrum characteristic of a Ca-rich area [middle spectrum of Fig. 5(e)] to the upper or lower PES spectrum of Fig. 5(e) with a scanning step of $4$ $\mu m$.
The abrupt changes in the PES spectra at the borders of two regions reveal that the experimental resolution is superior to this spatial step, thus extending to the nanometric domain.
Figs. 4-6 of this work provide evidence that the acquisition of ARPES images with a sub-micrometric spatial sensitivity is possible in the very same setup. This work has been focused on the STO(100) inhomogeneities in the micrometric scale. Different spatial probes have presented indications of Sr-rich \cite{Liang1994,Liang1995} and Ti-rich \cite{Szot1999,Newell2007} clusters at smaller spatial scales of thermally reduced STO(100). Further nano-(AR)PES measurements to identify the chemical nature of such inhomogeneities are in progress. 

\section{V. Summary}

We used Scanning Photoemission Microscopy for the direct imaging of Ca segregation sites on a STO(100) UHV-annealed single crystal surface. 
In the micrometric range, Ca impurities decorate surface defects by forming a ring of $15$ $\mu m$ around them.
Their segregation is strongly localized, has a negligible effect in the bulk and cannot be the reason of any long range reconstruction. The surface does not present any sign of Ca contamination outside the Ca-induced rings. 
Using an experimental tool which can provide (AR)PES spectra with spatial sensitivity up to the nanometric range, we examined the micrometric spatial modifications of the STO(100) electronic band structure 
in terms of the distribution of Ca impurities. In specific, the intragap electronic surface state is extinct in Ca-rich regions, while the bulk valence band shows variations in terms of signal-to-noise ratio. Our work
suggests that Ca contamination is a local phenomenon and
its effect on the electronic structure can be overcome with probes which combine high sensitivity in both real and reciprocal space.

\section{Acknowledgements}

The authors acknowledge the valuable support services of the Synchrotron SOLEIL and in particular the precious help of Stephane Lorcy, who has guaranteed a high quality technical assistance during the accomplishment of the experiments. This research was partially financed by PHC STAR projet N$^{\circ}$ 25852XH.

\end{document}